\documentclass[reprint,amsmath,amssymb,aps,pre,superscriptaddress]{revtex4-1}

\usepackage{dcolumn}
\usepackage{bm}

\usepackage{amsmath}
\usepackage{color}
\usepackage[usenames,dvipsnames,svgnames,table]{xcolor}
\usepackage[table]{xcolor}
\usepackage[english]{babel}
\usepackage{graphicx}
\usepackage{array}
\usepackage{multirow}
\usepackage{natbib}

\newcommand{\Tr}{\operatorname{Tr}}
\newcommand{\Matstyle}[1]{\textbf{#1}}
\newcommand{\avg}[1]{\ensuremath{\left<{#1}\right>}}

\begin{document}

\title{Evaluating functions of positive-definite matrices using colored noise thermostats}

\author{Marco Nava}
\affiliation
{Computational Science, Department of Chemistry and Applied Biosciences, 
ETH Zurich, USI-Campus, 
Via Giuseppe Buffi 13, C-6900 Lugano, Switzerland}

\author{Michele Ceriotti}
\affiliation{Laboratory of Computational Science and Modeling, {\'E}cole Polytechnique F{\'e}d{\'e}rale de Lausanne, 1015 Lausanne, Switzerland}

\author{Chaim Dryzun}
\affiliation
{Computational Science, Department of Chemistry and Applied Biosciences, 
ETH Zurich, USI-Campus, 
Via Giuseppe Buffi 13, C-6900 Lugano, Switzerland}

\author{Michele Parrinello}
\affiliation
{Computational Science, Department of Chemistry and Applied Biosciences, 
ETH Zurich, USI-Campus, 
Via Giuseppe Buffi 13, C-6900 Lugano, Switzerland}

\date{\today}

\begin{abstract}
Many applications in computational science require computing the elements of a 
function of a large matrix. A commonly used approach is based on the
the evaluation of the eigenvalue decomposition, a task that, in general, involves 
a computing time that scales with the cube of the size of the matrix.  
We present here a method that can be used to evaluate the elements of a function 
of a positive-definite matrix with a scaling that is linear for sparse matrices 
and quadratic in the general case.
This methodology is based on the properties of the dynamics of a multidimensional harmonic potential
coupled with colored noise generalized Langevin equation (GLE) thermostats. This
``$f-$thermostat'' (FTH) approach allows us to calculate directly 
elements of functions of a positive-definite matrix  
by carefully tailoring the properties 
of the stochastic dynamics. 
We demonstrate the scaling and the accuracy of this approach for both dense and sparse 
problems and compare the results with other established methodologies.
\newline
\end{abstract}

\maketitle

\section*{Introduction}

Eigenvalue equations are ubiquitous in science, as they play an 
essential role in quantum mechanics  \cite{bernardson+94cpc,kohn96prl,goed99rmp}, 
statistical mechanics  \cite{alav-frenk92jcp}, biology  \cite{porto+04prl}, graph
theory  \cite{bollobas+book}, computer science  \cite{bryan+06siam}, pure mathematics \cite{Bai+96jcam} 
and more \cite{bonacich+87ajs}. 

In many cases, one evaluates the eigenvalues as an intermediate step towards 
computing a function $f$ of the input matrix $\Matstyle{M}$.
In fact, the standard procedure to compute matrix functions is to 
diagonalize the input matrix $ \Matstyle{M} $ by finding its right
eigenvectors $ \mathbf{u}_i $ and its eigenvalues $\epsilon_i$. 
If $\Matstyle{U}$ is the 
matrix that has $ \mathbf{u}_i $ as columns and $\boldsymbol{\epsilon}=\mathrm{diag}\left\{\epsilon_i\right\}$ 
is a diagonal matrix with the eigenvalues on the diagonal, one can write
\begin{equation}
  \Matstyle{M}=  \Matstyle{U} \boldsymbol{\epsilon} \Matstyle{U}^{-1}, \quad\text{and}\quad
 f(  \Matstyle{M}) = \Matstyle{U} \mathrm{diag}\left\{f(\epsilon_i)\right\} \Matstyle{U}^{-1}.
\end{equation}
The most significant shortcoming of this approach is that conventional techniques for computing
the eigenvalue decomposition of a matrix of size $N$ have a computational cost that scales
with the cube of the size of the matrix, which makes them very inefficient when used 
on large matrices~\cite{golub89book}.

In many applications, however, one is dealing with very sparse matrices, and 
needs to compute just a small subset of the elements of $f(\Matstyle{M})$. 
In these cases, one should consider whether it is possible to obtain 
algorithms with a more favorable scaling. 
A typical example can be found  in the field of electronic structure calculations,
where the interaction potential between the particles can be obtained
by evaluating the density matrix $\boldsymbol{\rho}$ -- 
that at finite temperature is just the Fermi function
of the Hamiltonian matrix $\Matstyle{H}$ -- and then 
computing $\Tr \boldsymbol{\rho}(\Matstyle{H}) \Matstyle{H}$.

The efficiency of this kind of calculations for large systems can be improved greatly 
if one can evaluate $\boldsymbol{\rho}$ while avoiding the diagonalization of the 
Hamiltonian, using instead methods with a cost that scales linearly with the size of the system. 
Many different  linearly scaling algorithms have been suggested \cite{yang91prl,gall-parr92prl,li+93prb,goed-colo94prl,kohn96prl,pals-mano98prb,
goed99rmp,bowl-gill99cpc,kraj-parr05prb,kraj-parr06prb,kraj-parr06prb-2,kraj-parr07prb,kohn96prl,ceri+08jcp,ceri+09proc,lin+09prb}, but 
they are still not used routinely, in part because of their complexity and their memory requirements, and in
part because they only become advantageous for very large system sizes. 
In this paper we introduce a method based on colored-noise, generalized
Langevin equation (GLE) thermostats \cite{ceri+09prl,ceri+10jctc}, that can be used to compute selected 
elements of the function of a symmetric, positive-definite sparse matrix with
linear-scaling effort. 

Our approach is a generalization of the ideas underlying the so-called $\delta$-thermostat \cite{ceri-parr10pcs}.
By considering the input matrix as the Hessian of an $N$-dimensional harmonic oscillator,
one can generate an artificial, non-canonical GLE dynamics that selectively enforces different occupations 
onto the various normal modes. The key idea behind the $f$-thermostat is that 
the GLE thermostat can be tailored to generate an arbitrary frequency dependence of the 
fluctuations of position and momentum,
and that with an appropriate choice of such dependence one can obtain the elements
of a function of the input matrix as averages of the dynamical variables. 
We will start by describing the details of the methodology, and then present
its application to different symmetric matrices and real functions,
that will be used to discuss the advantages of the method, 
the statistic and systematic errors, and demonstrate the linear
scaling of its computational cost for large, sparse matrices.

\section{The f-thermostat}

A non-Markovian stochastic differential equation (SDE), based on the Langevin equation, 
for a particle with mass-scaled position $ q $ and mass-scaled momentum $ p $, moving 
under the influence of a potential $ V \left( q \right) $ can be written as   \cite{bern-fors71arpc,fox87jsp,abe+96prep,lucz05chaos}:
\begin{equation} 
\label{eq:nonmarkovian}
\begin{split}
  \dot{q} &= p  \\
  \dot{p} &= -\frac{\partial V\left ( q \right )}{\partial q} - \int {K\left ( t-s \right )  p\left ( s \right ) ds} + \zeta \left ( t \right ).
\end{split}
\end{equation}
Here $ K\left ( t \right ) $ is a memory kernel describing the friction, and 
$ \zeta \left ( t \right ) $ is a noisy-force term characterized by Gaussian statistics 
and time correlation function $ H\left ( t \right ) = \left<\zeta\left ( t \right )
\zeta\left ( 0 \right )\right> $. Treating analytically and integrating numerically
the SDE in this form is inconvenient. 
Generating a sequence of random numbers with the prescribed correlation
is possible but not straightforward~\cite{mara+90aop}. 
Furthermore, storing and referring to the past momenta in order 
to introduce the friction memory kernel is inefficient from the point of view of memory 
requirements and computation time. 

However, Eq. (\ref{eq:nonmarkovian}) can be represented \cite{zwan73jsp,zwan+01book,zwan61pr} by a 
Markovian, Langevin equation in an extended momentum space containing a certain number of additional 
momenta $\lbrace s_i \rbrace$. One can then write \cite{ceri+10jctc,ceri10phd},
\begin{equation}
\begin{split}
  \dot{q}=&p\\
\!\left(\! \begin{array}{c}\dot{p}\\ \dot{\mathbf{s}} \end{array}\!\right)\!=&
\left(\!\begin{array}{c}-V'(q)\\ \mathbf{0}\end{array}\!\!\right)
\!-\!\left(\!
\begin{array}{cc} 
a_{pp} & \mathbf{a}_p^T \\ 
\bar{\mathbf{a}}_p & \mathbf{A}
\end{array}\!\right)\!
\left(\!\begin{array}{c}p\\ \mathbf{s}\end{array}\!\right)\!+\!
\left(\!
\begin{array}{cc}
b_{pp} & \mathbf{b}_p^T \\ 
\bar{\mathbf{b}}_p & \mathbf{B}
\end{array}\!\right)\!
\left(\!\begin{array}{c}\multirow{2}{*}{$\boldsymbol{\xi}$}\\ \\\end{array}\!\right),
\end{split}
\label{eq:mark-sde}
\end{equation}
where $\boldsymbol{\xi} $  is a vector of $n+1$ uncorrelated Gaussian 
random numbers satisfying
$ \left< \xi_i \left(  t \right ) \xi_j \left(  {t'} \right ) \right> = \delta_{ij}  
\delta \left(  t-t' \right ) $ and $\langle\xi_i\rangle=0$. 
By integrating out the additional fictitious degrees of freedom, the non-Markovian 
Langevin equation~\eqref{eq:nonmarkovian} is recovered describing the trajectory of the physical 
degrees of freedom $ \left ( q,p  \right ) $, with memory kernels that correspond 
to a fairly general superimposition of complex exponentials \cite{ceri+10jctc,ceri10phd}.
In this Markovian form, the dynamics is determined by the drift and diffusion matrices
\begin{equation}\label{cpeq2}
\mathbf{A}_p=
\left(\!
\begin{array}{cc} 
a_{pp} & \mathbf{a}_p^T \\ 
\bar{\mathbf{a}}_p & \mathbf{A}
\end{array}\!\right) \quad\text{and}\quad
\mathbf{B}_p=
\left(\!
\begin{array}{cc}
b_{pp} & \mathbf{b}_p^T \\ 
\bar{\mathbf{b}}_p & \mathbf{B}
\end{array}\!\right)
\end{equation}
respectively. It is often more convenient to characterize the dynamics in terms
of the free-particle static covariance $\mathbf{C}_p=\avg{(p,\mathbf{s})(p,\mathbf{s})^T}$, i.e. the matrix that
contains the stationary correlations between $p$ and $\mathbf{s}$ in the absence
of an external potential.  
$\mathbf{C}_p$ is related to $\mathbf{B}_p$ and $\mathbf{A}_p$ through
\begin{equation}\label{cpeq}
\mathbf{A}_p \mathbf{C}_p + \mathbf{C}_p
\mathbf{A}_P^T =
\mathbf{B}_p\mathbf{B}_p^T,
\end{equation}
and can be used to express a number of physical constraints on the dynamics
and to derive a relatively straightforward integrator for Eqs.~\eqref{eq:mark-sde}, 
based on a modified velocity-Verlet algorithm~\cite{ceri+10jctc,ceri10phd}.

What makes the formalism of Eq.~\eqref{eq:mark-sde} particularly useful is that
the sampling and dynamical properties of the GLE trajectory can be predicted 
analytically in the case of a harmonic potential $V(q)=\omega^2 q^2/2$. 
This made it possible to obtain within the same formalism 
a number of useful effects, ranging from efficient canonical sampling~\cite{ceri+09prl,ceri+10jcp} 
to the stabilization of multiple time step dynamics~\cite{morr+11jcp} and
the modeling of nuclear quantum effects~\cite{ceri+09prl2,ceri+11jcp,ceri-mano12prl}.

Within this framework, one can obtain generalized Langevin dynamics that do not fulfill
the classical fluctuation-dissipation theorem, and therefore do not sample the 
canonical ensemble at a given temperature. In the harmonic limit, this stochastic
dynamics can be understood in terms of a steady-state process in which normal modes
of different frequencies reach a stationary distribution in which position and momentum
have Gaussian statistics, with frequency-dependent fluctuations. 
Manipulating the behavior of $\avg{q^2}(\omega)$ and $\avg{p^2}(\omega)$ is the 
basic idea behind the ``quantum thermostat''\cite{ceri+09prl2}, the 
$\delta$-thermostat~\cite{ceri-parr10pcs}, and the method we introduce in this paper.
Indeed, the $f$-thermostat can be viewed as a generalization of the $\delta$-thermostat,
which uses non-canonical sampling to thermalize at a higher 
effective temperature only the characteristic frequencies of the system that lie in a 
prescribed, narrow frequency range. 
The other frequencies equilibrate at a much lower effective temperature. This is done by enforcing 
a delta-like frequency dependence of the fluctuations through a GLE based thermostat, and 
can be used to estimate the density of states of a positive-definite matrix~\cite{ceri-parr10pcs}.
The idea behind the present method is that by a careful choice of the frequency
dependence of the fluctuations one can directly evaluate the elements of $f(\Matstyle{M})$.

To see how this can be achieved, let us consider a symmetric, positive-definite matrix, $ \Matstyle{M} $ 
of size $N$. This matrix can be used as the Hessian of a multi-dimensional harmonic 
potential:
\begin{equation} 
\label{eq:quad-v}
V\left ( \textbf{q} \right ) = \frac{1}{2} {\textbf{q}}^{T} {\Matstyle{M}} {{\textbf{q}}}.
\end{equation}
One could then study the dynamics of a particle with unit mass subject  
to this harmonic potential. The resulting problem is best studied in terms of
normal mode coordinates: if the $k$-th eigenvalue of $\Matstyle{M}$ is $\epsilon_k$, 
and the associated eigenvector $\mathbf{u}_k$, one can write the displacement along
the normal mode as $\tilde{q}_k = \mathbf{u}_k^T\mathbf{q}$ and 
the corresponding momentum as $\tilde{p}_k = \mathbf{u}_k^T\mathbf{p}$.
If the dynamics is microcanonical, this normal mode will oscillate indefinitely
with a frequency $ \omega_{k} = \sqrt{ \epsilon_{k}} $.  If instead
the system is coupled to a heat bath at temperature $T$ -- for 
instance by using a Langevin equation -- the fluctuations of 
displacement and momentum are bound to be $\avg{\tilde{q}_k^2}=T/\omega_k^2$ and 
$\avg{\tilde{p}_k^2}=T$, taking for simplicity $k_B=1$. 

Now let us consider a GLE thermostat whose parameters have been fitted in such 
a way that, over the range of frequencies that corresponds to the spectral range
of $\Matstyle{M}$, a harmonic oscillator of frequency $\omega$ will have momentum 
fluctuations $\avg{p^2}(\omega)=T^\star(\omega)$, where we introduced the effective
temperature $T^\star$ as a shorthand for the average momentum fluctuations.
Let us assume that the GLE parameters have been chosen so that the momentum fluctuations
are related to the desired matrix function $f$ as $T^\star(\omega)=f\left(\omega^2\right)$.
If we then apply an independent thermostat of this kind to each of the $N$ ``Cartesian'' degrees of 
freedom of the artificial dynamics generated by the potential~\eqref{eq:quad-v}, each normal mode
will respond as if the GLEs were applied in the normal modes representation. 
This is a consequence of the invariance of the combined GLE dynamics with respect 
to a unitary transformation of the coordinates, which in turn is guaranteed by the Gaussian
nature of the noisy force and by the linear nature of the stochastic differential equations~\cite{ceri10phd}.
The individual normal modes described by the eigenvalues of $\Matstyle{M}$ will then respond to 
the GLE dynamics and reach a steady state with fluctuations that are consistent with
the thermostat specifications; namely 
$\avg{\tilde{p}_k\tilde{p}_{k'}}=\delta_{kk'}T^\star(\omega_k)=\delta_{kk'}f(\epsilon_k)$. 
Since the 
eigenvectors $u_k$ form an orthonormal basis and $p_i = \sum_k \tilde{p}_k U_{ik} $, 
one can write 
\begin{eqnarray}
\avg{p_i p_j} = \sum_{kk'} U_{ik}U_{jk'}\avg{\tilde{p}_k \tilde{p}_{k'}}= \nonumber \\
= \sum_k U_{ik} f(\epsilon_k) U_{jk} = f(\Matstyle{M})_{ij}.
\label{eq:pipj}
\end{eqnarray}
In particular, $\Tr f(\Matstyle{M})$ can be obtained 
very easily from the kinetic energy of the GLE-thermostatted harmonic dynamics:
$\Tr f(\Matstyle{M})=\sum_i \avg{p_i^2}=\avg{\mathbf{p}^T\mathbf{p}}$. 

Provided that $\Matstyle{M}$ is sparse, the technique described above can be 
applied with an effort that scales linearly with the size of the matrix, 
because computing the forces which correspond to the potential~\eqref{eq:quad-v} involves a 
matrix-vector product that can be evaluated in $\mathcal{O}(N)$. Averages
that correspond to individual elements of the momentum covariance matrix 
can be evaluated with an effort independent of the size of $\Matstyle{M}$,
so the method remains linear scaling as long as one needs to compute
a $\mathcal{O}(N)$ subset of the elements of $f(\Matstyle{M})$.
As we will discuss later, the relative statistical error on the results 
is also independent of $N$. 
The procedure we have highlighted here allows to obtain elements of $f(\Matstyle{M})$ for large, sparse,
positive-definite matrices without performing explicitly an eigenvalue decomposition,
since both the 
energy~\eqref{eq:quad-v} and the momentum covariance~\eqref{eq:pipj} can be computed
in the ``Cartesian'' representation. 
In a similar way one can also obtain traces of the form $\Tr \Matstyle{O} f(\Matstyle{M})$, 
which often appear when computing observables in electronic structure calculations and
when evaluating thermal averages. In fact, it is easy to see that $\Tr \Matstyle{O} f(\Matstyle{M})= 
\avg{\Tr \Matstyle{O}\mathbf{p}\mathbf{p}^T}= \avg{\mathbf{p}^T\Matstyle{O}\mathbf{p}}$. 

While the idea behind this approach is relatively simple, it is also true that designing a thermostat that
enforces the desired $\avg{p^2}(\omega)$ dependence is not trivial \cite{ceri+10jctc,ceri10phd}. 
Many conflicting goals must be fulfilled besides the prescribed $\omega$-dependent momentum
fluctuations, e.g. the stability of the dynamics and the efficiency of the sampling. 
The fitting strategy we use is based on that described in 
Refs.~\cite{ceri+10jctc,ceri10phd}, but we found advantageous to start with a simplified
parametrization of the $\Matstyle{A}_p$ and $\Matstyle{C}_p$ matrices that corresponds to
a combination of $\delta$-like memory kernels~\cite{ceri10phd,ceri-parr10pcs}. 
The resulting matrices can then be fine-tuned using the more general parametrization
described in Ref.~\cite{ceri+10jctc}.
In practice, obtaining an effective set of parameters requires some experience and
a few hours of work. However, as discussed in previous publications, once a pair of 
$\mathbf{A}_p$ and $\mathbf{C}_p$ matrices has been obtained for a given application,
it is possible to re-use them in many similar circumstances, by modifying the
range of frequencies they work on just by scaling their value.

In the case of the $f$-thermostat, scaling the GLE parameters can also be used to
extend the range of functions that can be computed. In fact, for the
artificial dynamics to yield a well-defined stationary distribution
 $\Matstyle{M}$ must be positive-definite, and 
$f$ must be positive-valued, so that the method is restricted to the evaluation of 
$f:D\subseteq\mathbb{R}^+\rightarrow C\subseteq\mathbb{R}^+$. 
Shifting and scaling $\mathbf{A}_p$, $\mathbf{C}_p$ and $\Matstyle{M}$ can be used 
to change the spectral range of the input matrix to make it positive-definite,
while still obtaining the desired matrix function.

Let us consider the scaling of the matrices in Eq. (\ref{cpeq}) and (\ref{cpeq2}), 
$\Matstyle{A}'_p=\alpha \Matstyle{A}_p$ and
$\Matstyle{C}'_p=\gamma \Matstyle{C}_p$, and the transformation of $\Matstyle{M}$ into 
$\Matstyle{M}'=r\Matstyle{M}+m\Matstyle{I}$ -- where $\Matstyle{I}$ is the
$N\times N$  identity matrix.
If $\Matstyle{M}$ has eigenvalues within the range 
$\left[\epsilon_\textrm{min}, \epsilon_\textrm{max}\right]$, then one can obtain 
a transformed matrix with range $\left[\epsilon'_\textrm{min}, \epsilon'_\textrm{max}\right]$
by choosing 
$r=\left(\epsilon'_\textrm{min}-\epsilon'_\textrm{max}\right)/\left(\epsilon_\textrm{min}-\epsilon_\textrm{max}\right)$ 
and $m=\left(\epsilon_\textrm{max}\epsilon'_\textrm{min}-\epsilon_\textrm{min}\epsilon'_\textrm{max}\right)/\left(\epsilon_\textrm{min}-\epsilon_\textrm{max}\right)$. 
In the most general case, when all the transformations are applied simultaneously, the $f$-thermostat  
procedure would yield $\avg{\mathbf{p}\mathbf{p}^T}=f'(\Matstyle{M}')=\gamma f\left((r\Matstyle{M}+m \Matstyle{I})/\alpha^2\right)$.
Depending on the function $f$ it might still be possible to perform the inverse transformation and recover the original target.  
For instance, if $f(x)\equiv \ln(x)$, then  $f\left(\Matstyle{M}\right)=f\left(r\Matstyle{M}\right)-\Matstyle{I}\cdot \ln r$,
and if $f(x)\equiv \exp(x)$, then  $f\left(\Matstyle{M}\right)=f\left(\Matstyle{M}+m\Matstyle{I}\right)e^{-m}$,

Let us stress from the start that this does not aim to be a very accurate method.
A first source of error is the fact that the fit will not in general be 
capable of enforcing perfectly the desired behavior for $\avg{p^2}\left(\omega\right)$. 
More flexibility in the fit -- and a tighter compliance to the desired constraints -- can 
be achieved to an extent by increasing the number $n$ of additional degrees of freedom, 
at the cost however of larger memory and computational requirements. 
Typically, we find that a choice between $4$  and $10$ additional momenta $\lbrace s_i \rbrace$ 
provides sufficient flexibility to obtain a maximum relative error
of one percent or less over a range of frequencies spanning several orders of magnitude. 
Other errors can originate from the parameters of the simulation 
and can be in principle be made arbitrarily small. 
For instance, finite time-step errors can be minimized by choosing small enough time steps; 
when using a velocity Verlet integrator~\cite{ceri+10jctc}, one could also perform the fit
for a specific integration time-step, modifying the target fluctuations vs. frequency curve in such a way 
that finite time-step errors are accounted for and exactly corrected.
The most serious limitation of this method, however lies in its stochastic nature.
The elements of $f(\Matstyle{M})$ are obtained from an average of dynamical variables,
and the average will be affected by a statistical error. This component of the error 
decreases slowly with simulation time, specifically with the square root of the number 
of uncorrelated samples. The method we propose here is best suited for 
the (many) applications in which a non-negligible  stochastic error can be tolerated 
and, perhaps, even useful~\cite{kraj-parr05prb,kraj-parr06prb}, or to initialize other 
approaches that converge faster but require an initial guess for $f(\Matstyle{M})$.

\section{Results and benchmarks}

As discussed, the $f$-thermostat relies on the accuracy of the fit of the frequency-dependent
fluctuations to the desired target; however 
the quality of the results as well as the performance of the methodology depend also on other 
constraints that we will discuss here briefly. In summary, a successful set of colored-noise parameters should guarantee:
(1) a tight agreement between $T^\star(\omega)$ and $f(\omega^2)$; (2) stability of the dynamics
with respect to the integration time step; (3) statistical efficiency. 

The first requirement is rather obvious since the discrepancy in the fit of  $T^\star(\omega)$ 
will cause systematic errors in the evaluated matrix function. It is not hard
to obtain small relative errors over a large range of frequencies for slowly-varying functions. 
However, the frequency-dependence of $T^\star(\omega)$ for simple forms of the GLE matrices have 
typically an algebraic $\propto 1/\omega^2$ decay for high values of $\omega$. This is particularly
problematic when one wants to obtain an exponentially-decaying $T^\star(\omega)$.
In many applications it is sufficient that the small $f(\epsilon)$ are negligible with
respect to the larger ones. Therefore, when fitting exponential tails,
we have decided to choose a small threshold $\delta$, and just require that for 
large $\omega$ the effective temperature of normal modes is smaller than 
$\delta \max \left[T^\star(\omega)\right] $ -- rather than it matches the exponential decay. 

The second requirement has an impact on the efficiency of the simulation. The 
time discretization $dt$ of the dynamics in the absence of noise is determined by the largest
eigenvalue of $\Matstyle{M}$, i.e.  $dt \ll 1/\sqrt{\epsilon_{\rm max}}$. However, if the noise
contains strong components with $\omega>\sqrt{\epsilon_{\rm max}}$, it may become necessary to 
reduce the time step further, in order to integrate properly the stochastic dynamics and obtain
the fluctuations that are predicted analytically for an infinitesimal time step. 
In practice, this constraint corresponds to making sure that the Fourier transform 
of the memory kernels $K(\omega)$ and $H(\omega)$ do not exhibit prominent features 
for $\omega>\sqrt{\epsilon_{\rm max}}$, and that 
$\lim_{\omega\rightarrow\infty}K(\omega)<\sqrt{\epsilon_{\rm max}}$. 
These requirements can be implemented by including additional penalty terms in the fitness function,
as discussed in Ref.~\cite{ceri+10jctc}.

The third requirement determines how quickly the stochastic dynamics generates 
uncorrelated samples of positions and momenta, and is therefore essential to speed up 
the convergence of averages with time. Since the method requires the calculation
of $\avg{p_i p_j}$, it is important to minimize the correlation time of $p^2$,
$\tau_{p^2}(\omega)$, over the range of frequencies that is relevant for the problem at hand. 
If $t_{\rm tot}$ is the total simulation time, the statistical error on the value of 
$\avg{p^2}$ for a normal mode of frequency $\omega$ will be of the order
of $\sqrt{\tau_{p^2}(\omega)/t_{\rm tot}}$ -- so minimizing  $\tau_{p^2}(\omega)$ allows one to obtain
smaller statistical uncertainty for a given length of the simulation. 
A small autocorrelation time is also necessary to make sure that the final result will be independent on the initial
conditions of the dynamics -- which could in principle introduce a systematic bias in the estimate of $f(\Matstyle{M})$.

A remarkable feature of the FTH method is the very low memory consumption.
If $\Matstyle{M}$ is sparse, the memory footprint grows linearly with the size of the matrix:
the storage required consists of roughly $M + (3+S)N + 2A$ real numbers. 
The first term is required for the storage of matrix $\Matstyle{M}$ in sparse format -- corresponding to the number $M$
of non-zero entries;  the second term is the memory required for the modified Verlet algorithm, where $N$ is the size of $\Matstyle{M}$
and $S$ is the number of additional momenta, and the last term is the storage required to collect and block-average an arbitrary number $A$ of
elements of $f(\Matstyle{M})$ that one wishes to compute. The memory required by the FTH method in the case of a dense matrix is also relatively small 
and grows quadratically with the size of $\Matstyle{M}$: one would need $N(N+1)/2$ elements to store $\Matstyle{M}$, another $N(N+1)$ for the 
block-averaging of the entire  $f(\Matstyle{M})$ and a linear-scaling $(3+S) N$ for the propagation of the GLE dynamics.

Let us now present a few examples, where we apply the $f$-thermostat to compute 
various matrix functions for matrices of different form.

\begin{figure}[t]
\includegraphics*[scale=0.33]{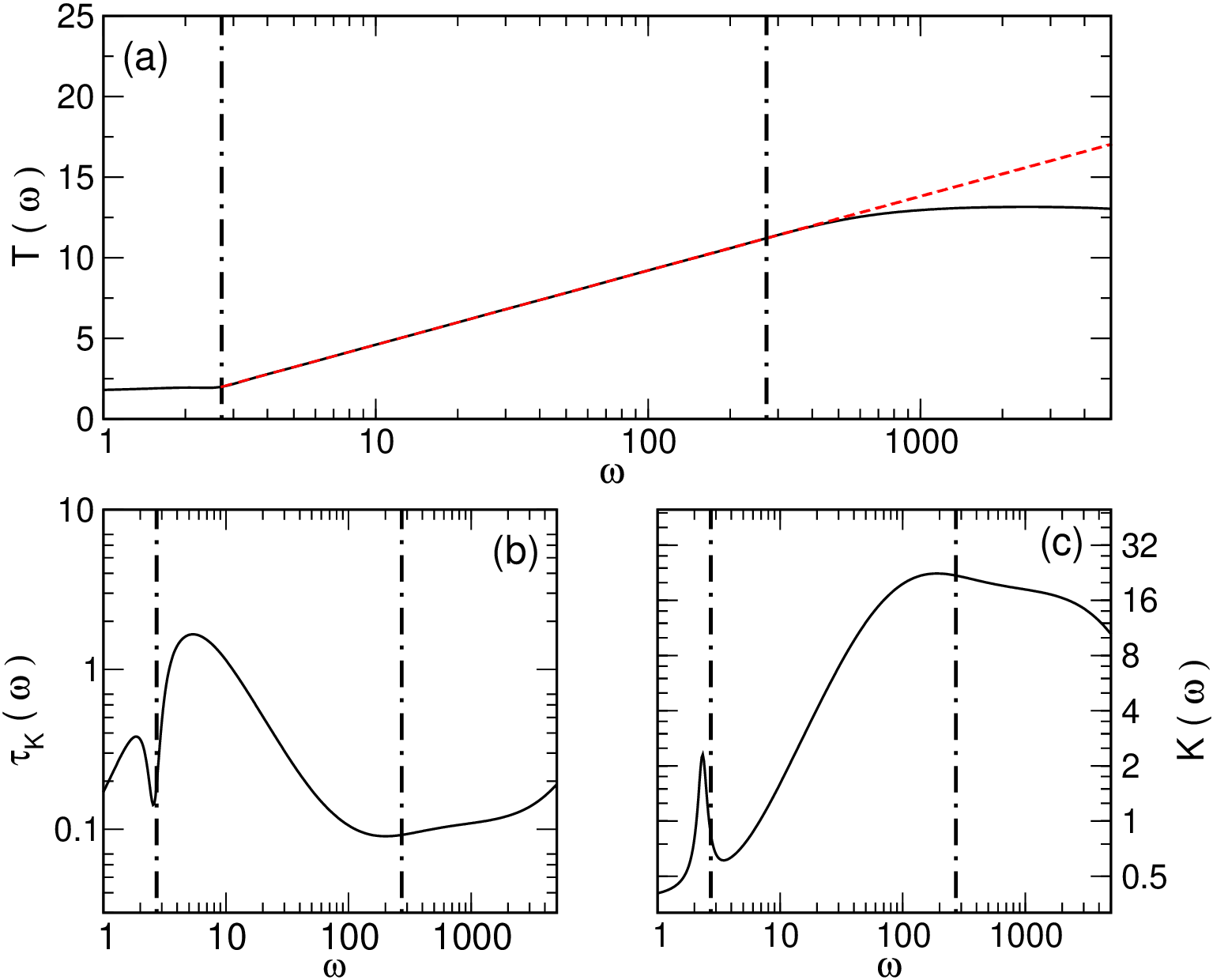}
\caption{(Color online) 
The fit of the fluctuations versus frequency dependence of the thermostat along with the 
constraints required for an efficient thermostatting of the system. (a) 
$\avg{p^2}$ dependency and (Dashed line) the fit target, the fit interval is marked by the vertical dotted lines; 
(b) Autocorrelation time of $p^2$ 
as a function of frequency; (c) Effective friction -- the Fourier transform of the memory kernel $K(\omega)$. 
Arbitrary units have been used. 
}
\label{fig1}
\end{figure}

\begin{figure}[t]
\includegraphics*[scale=0.31]{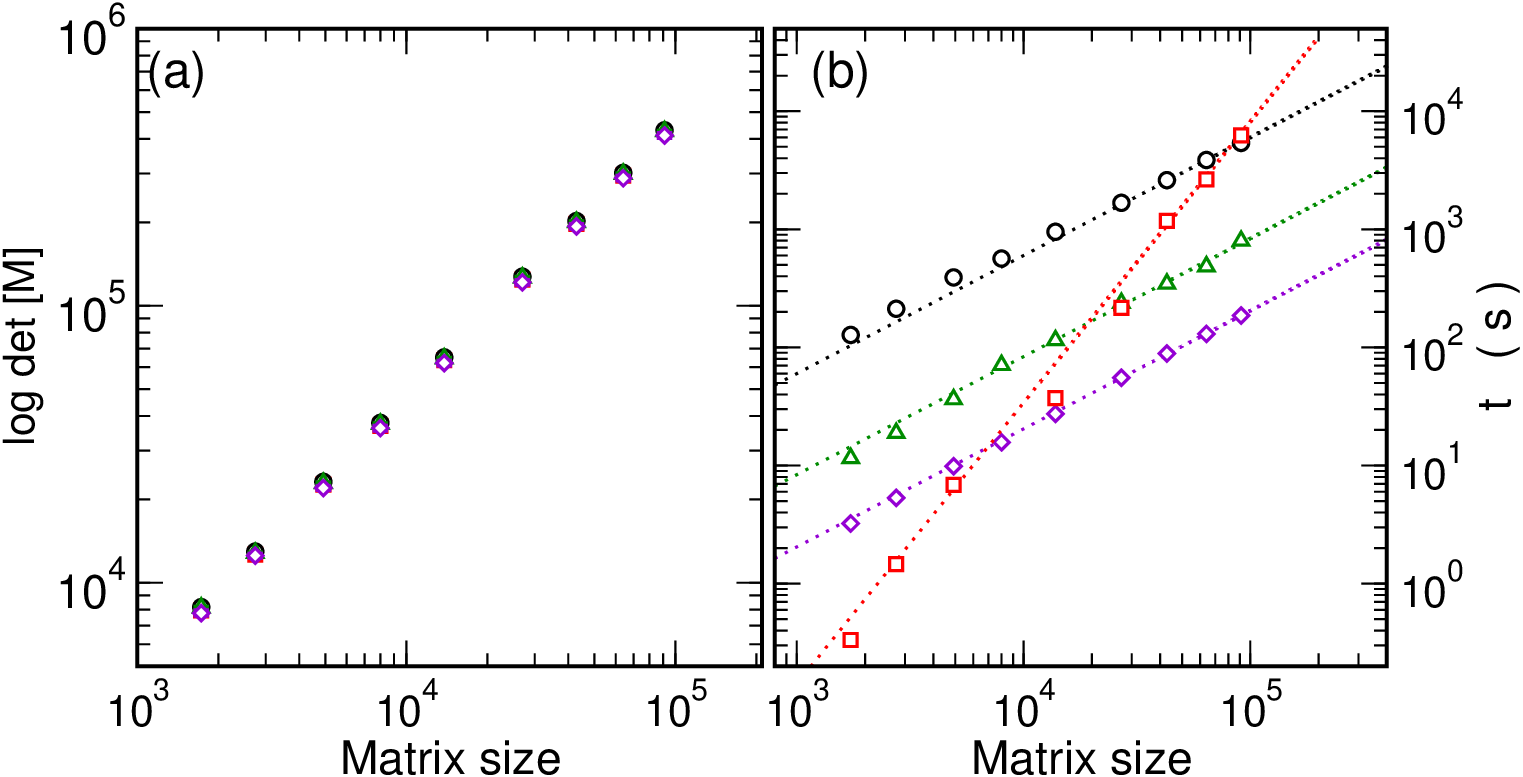}
\caption{(Color online) 
Panel a) logarithm of the determinant of a sparse matrix $\Matstyle{M}$ obtained from the trace of the 
logarithm of $\Matstyle{M}$ with a Sparse Cholesky decomposition (Squares) and 
the $f$-thermostat approach with a 0.1\% (Circles), 1\% (Triangles) and 2.5\% (Diamonds) uncertainty on the results.
Panel b) The computing time required to evaluate the quantities displayed in panel (a). Symbols 
have the same meaning as those in panel (a). Dotted lines are power-law fits of the computing time.
}
\label{fig2}
\end{figure}

\paragraph*{Matrix logarithm.}

We chose as a first example the problem of evaluating the logarithm of a matrix.
Fig.~\ref{fig1} shows the $\avg{p^2}(\omega)$ curve for an optimized set of parameters, 
following the desired $\ln \omega^2$ behavior for frequencies within the interval
$\left[e,100 e\right]$. 
The figure also shows the sampling efficiency and the friction kernel, whose 
importance has been discussed in the previous paragraph.

In Fig.~\ref{fig2} we show the values obtained for the evaluation of the trace of the 
logarithm of matrices $\Matstyle{M}_1$ of different size, using the colored noise 
parameters described in Fig.~\ref{fig1}, together with the wall-clock time needed to 
converge the results with different levels of accuracy. The results and timings are compared
with those obtained  with a supernodal sparse Cholesky decomposition\cite{cholmod} 
of $\Matstyle{M}_1$, an highly optimized tool for the computation of determinants 
of symmetric positive definite matrices that, through the relation 
$\ln\det(\Matstyle{M}) = \Tr \ln \Matstyle{M}$, allows for a check of the results.
Here $\Matstyle{M}_1$ matrices 
have the form of the tight binding Hamiltonian of a three dimensional cubic lattice 
with periodic boundary conditions with up to the sixth nearest neighbors random hopping 
terms and a constant on-site interaction that keeps the matrix positive definite. 
Details on how $\Matstyle{M}_1$ matrices have been defined are reported in the Appendix.

Matrices of this form are frequently encountered in physics and it is known that, in general, 
it is not possible to transform them into banded matrices that can be manipulated 
with linear scaling effort by sparse linear algebra methods. In fact, 
the time required for the supernodal Cholesky decomposition (SC) scales super-quadratically 
with matrix size, as $\mathcal{O}(N^{2.4})$. As expected, the FTH method scales 
linearly with system size. For small matrices the SC method is faster -- SC is
a highly optimized method specifically aimed at computing a matrix determinant, so 
a direct comparison with the more general-purpose FTH is somewhat unfair -- but
its worse system-size scaling will inevitably lead to the FTH method to prevail for large
enough $N$. As it is apparent from Fig.~\ref{fig1}, the break-even point depends 
dramatically on the desired level of statistical accuracy. For a relative error
of the order of 0.1\%, the break-even point is at  $N \sim 10^5$.

\begin{figure}[t]
\includegraphics*[scale=0.31]{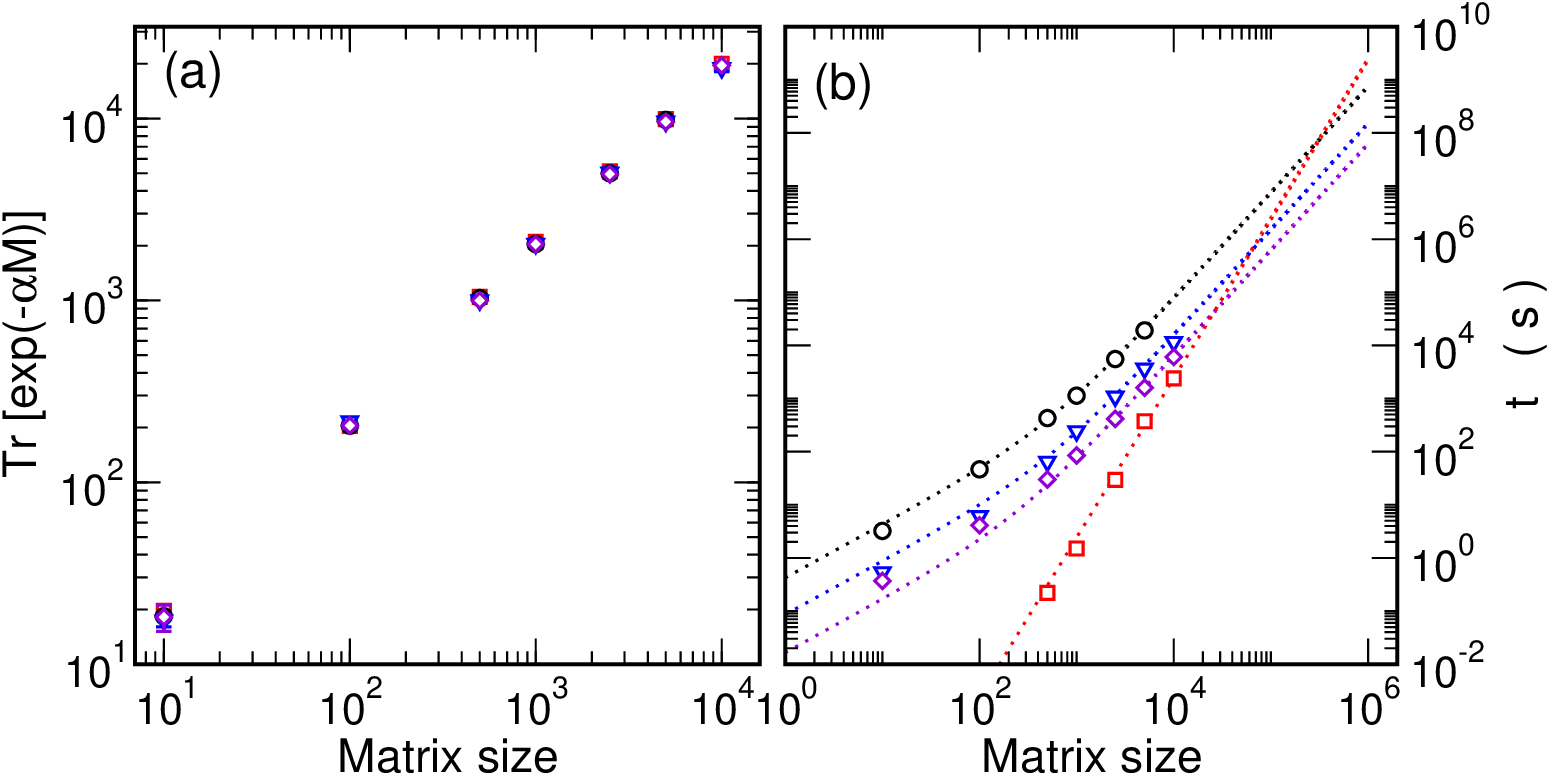}
\caption{(Color online) 
Panel a) (Squares) $\Tr[\exp(-\alpha \Matstyle{M})]$ obtained with a diagonalization approach~\cite{eigen}. Other symbols: the same obtained with the $f$-thermostat approach with 
uncertainties of 0.2\% (Circles), 0.5\% (Triangles) and 1\% (Diamonds) on the results.
Panel b) Time scaling for the computations in panel (a). Symbols have the same meaning of those in panel (a). Dotted lines are fit to computing time data. 
}
\label{fig3}
\end{figure}


\paragraph*{Matrix exponential.}
The FTH methodology can also be used in the case of dense matrices, where its computational cost grows as $\mathcal{O}(N^2)$. 
Fig. \ref{fig3} shows a comparison between the FTH method and 
a diagonalization method \cite{eigen} for the computation of the exponential of a dense symmetric random matrix $\Matstyle{M}_2$ 
with eigenvalues in the interval $(a,b)\approx(-300,300)$; namely $\exp(-\alpha \Matstyle{M}_2)$ with $\alpha = 1/100$ (see the Appendix for more
 details on the form of $\mathbf{M}_2$). 

The matrix exponential is a typical operation encountered in many fields of physics --  
for instance when computing thermal density matrices $\exp(-\beta\hat{H})$. For this operation, 
the conventional diagonalization approach scales with the cube of the matrix size, so again for large enough matrices
the FTH method will prevail. Note that for small $N$ the cost of the FTH method grows linearly with matrix size, as it is dominated by the 
(linear scaling) cost of propagating the stochastic dynamics and not by the evaluation of the force. 


Comparing the timings reported in Fig. \ref{fig1} and Fig. \ref{fig3} in the size range in which the 
propagation of the dynamics dominates the computational cost of the FTH, one can see that for a given accuracy
evaluating $\exp(\Matstyle{M})$ is much more demanding than evaluating $\ln(\Matstyle{M})$. 
This is a consequence of the fact that the autocorrelation time $\tau_{p^2}$  we could obtain for the exponential
fit was longer than in the case of the logarithm -- so a longer artificial dynamics is necessary to obtain
the same statistical uncertainty. This in turn is a consequence of the difficulty of obtaining the sharp 
decay which is characteristic of the exponential function, as explained above. 
A better efficiency could be obtained by using a larger threshold for the exponential tail (here we 
tried to fit the exponential decay down to $e^{-x}\approx 10^{-4}$), or by targeting specifically the range of frequencies
that is relevant to the application at hand. 


\paragraph*{Fermi function.}
The examples reported this far had the purpose of benchmarking the methodology and did not have
any direct physical relevance. We show now an application of the FTH method to a concrete physical 
problem -- namely, the evaluation of the electrons per site $\rho_s$ and the energy $U/N_{at}$ 
for a cubic supercell containing $N_{at}$ carbon atoms in a diamond lattice, with lattice constant $l=3.57$~\AA,  
and at zero temperature, described by a tight binding Hamiltonian\cite{carbontb} with only first and second nearest neighbors interactions.

\begin{figure}[t]
\includegraphics*[scale=0.32]{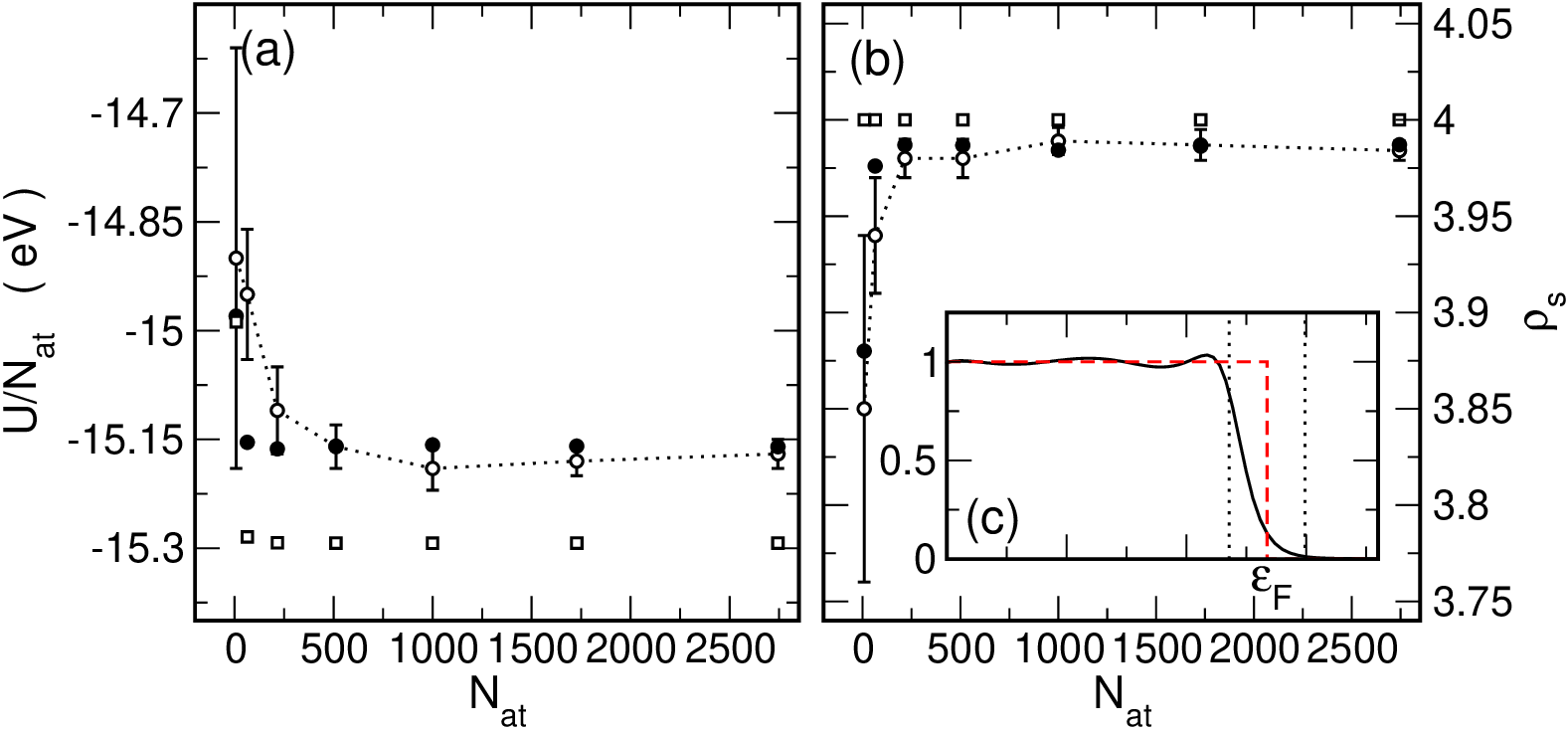}
\caption{ 
(a) Internal electron energy per atom and (b) average electron number per atom of a diamond bulk carbon lattice 
with lattice constant $l=3.57$~\AA{} and temperature $T\rightarrow 0$~K. 
The Fermi energy level has been placed in the middle of the energy gap so that $N_e/N =4$. Open circles refer to 
the results obtained with the $f$-thermostat method, closed circles are the same data obtained by exact diagonalization
but computing the approximate fit to the Fermi function (i.e. they correspond to the result predicted for the GLE 
dynamics in the absence of time step and statistical error).  Open squares are the exact results computed
from the eigenvalues using a step function.
(c) (full line) The approximate Fermi function obtained from the fluctuation profile $\langle p^2\rangle (\omega)$ of the $f$-thermostat
we used for this example; 
the dashed (red) line marks the zero-temperature Fermi function and the vertical dotted lines show the boundaries of the energy gap. 
}
\label{fig4}
\end{figure}
Given the tight binding Hamiltonian matrix $H$ that describes the valence electrons of carbon, $U/N_{at}$ and $\rho_s$ can be computed 
with occupations given by the Fermi function $f_F(\epsilon)$,
\begin{eqnarray}
f_F(\epsilon)=\frac{1}{e^{\beta(\epsilon-\epsilon_F )}+1}\\
\rho_s = 2\Tr \left[ f_F(H)\right] \label{eqthf1}/N_e \\
U/N_{at} = 2\Tr \left[ H\cdot f_F(H) \label{eqthf2}\right]/N_{at}
\end{eqnarray}
where the factor of 2 takes into account the spin degeneracy of the electron,  
$\epsilon_F$ is the Fermi energy of the system and $N_e$ is the number of electrons,

In the low-temperature limit,  
$f_F(\epsilon)\rightarrow \theta(\epsilon_F-\epsilon)$ (a Heaviside theta function), and would be 
very difficult to reproduce with a $f$-thermostat, because one would have to fit $T^\star(\omega)$
to a sharp, discontinuous target.
In the case of an insulator, however, it is not necessary to reproduce precisely $f_F$, since $\epsilon_F$ falls 
in the energy gap and thus a smooth transition of $T^\star(\omega)$ between 0 and 1 does not affect results, as long as
it is confined to a neighborhood of $\epsilon_F$ entirely contained in the gap. 
This is the case shown in Fig. \ref{fig4}c; the energy gap for a diamond lattice of carbon atoms, 
in fact, is roughly 6.5 eV and for the computation of properties that do not involve a
shift in the Fermi level, $\epsilon_F$ can be safely placed in the middle of the gap; 
the frequency-dependent fluctuations have been fitted 
with a maximum error outside the gap of $3$\%, the most relevant discrepancy being the oscillation shortly before the energy gap.
In the case of the carbon Hamiltonian, this introduces a systematic error of 
about 1\% on the values of $\rho_s$ and $U$, that is clearly visible in Fig. \ref{fig4}b.
Let us stress that if one needed to perform a self-consistency procedure, or to modify $\epsilon_F$ to enforce
constant electron number when considering different structures as it would be the case when optimizing
the geometry or running a molecular dynamics simulation, it would be possible to just scale the $\Matstyle{A}_p$
and/or $\mathbf{M}$ matrices without having to perform a new fit for each configuration.

All data in Fig. \ref{fig4}a and \ref{fig4}b has been obtained with the same parameters for the Verlet dynamics. 
The relative statistical error is stable with respect to the size of the matrix, which is clearly a desirable feature
of a method aimed at large sparse matrix calculations. Note in passing that the $f$-thermostat does not require 
an initial guess for $f(\Matstyle{M})$, contrary to other methods to compute matrix functions in general, 
and the density matrix in particular. For instance, a widely used approach to compute the zero-temperature
Fermi function is to perform Newton--Schulz iterations \cite{funmatrix} that converge to the sign function. 
This procedure has a convergent behavior only if the starting point is already a good estimate of $f(\Matstyle{M})$. 
In this and similar cases the $f$-thermostat can provide a sufficiently accurate estimate of $f(\Matstyle{M})$ to 
be used as starting point of the iteration.

\section{Final discussion}
In this work we have used the properties of a generalized Langevin dynamics coupled to a multi dimensional harmonic 
oscillator in order to compute the elements of a function of a positive-definite matrix.
We have verified that this approach entails a linear-scaling computational cost when the input matrix is sparse,
and a quadratic cost in the case of a dense input matrix. It is therefore an efficient method
when one needs to compute a function of a very large matrix. 

Tuning the parameters of the GLE thermostat is a very delicate step. The frequency-dependent effective temperature
$T^\star(\omega)$ has to be compatible with the target function in the relevant frequency range; at the same time,
the strength of the noise and the autocorrelation time of $p^2$ should be minimized to guarantee
stable dynamics and ergodic sampling.
Discrepancies in the fit introduce systematic errors, whereas the other constraints 
affect also the performances of the computation. In particular, a small autocorrelation time 
reduces the statistical error for a given length of the simulation.

The stochastic nature of the FTH method implies that the result will be affected by a slowly-decaying 
statistical error -- the simulation time must be multiplied by four just to half the statistical uncertainty. 
There is a trade-off between accuracy and computational cost, that in practice will make 
this idea applicable only in cases where a relatively large error, of about 1\%, can be tolerated -- 
for instance when the results are to be used as the forces in a Langevin dynamics, or as the initial
guess for a more accurate iterative method.
This said, the FTH method offers three main advantages: on top of the favorable scaling of computational time 
with the size of the input matrix, one should also consider the very small memory footprint, and the ease of
parallelization -- not only the matrix-vector multiplications can be parallelized, but one could perform independent,
trivially-parallel simulations to reduce the statistical error.

\appendix
\section{Details on the form of the example matrices\label{appe}}
Here we report the prescription for generating the $\mathbf{M}_1$ and $\mathbf{M}_2$
matrices that we used in our examples. The parameters of the $f$-thermostat that we used
are provided in the Supporting Information \cite{SI}, together with a simple program that
performs the artificial GLE dynamics and computes the elements of $f(\Matstyle{M})$.
\paragraph*{$\mathbf{M}_1$ matrices (logarithm example).}
The on-site terms have been set to $U=100$ (arbitrary units) and the hopping terms 
are random numbers between $0$ and $2$. Having fixed a number $n$ of sites per directions of the lattice, 
the diagonal elements $(i,i)$ correspond to the on-site terms, whereas 
the off-diagonal elements $(i,j)$ of the matrix correspond to the
hopping terms between the elements $(i_x,i_y,i_z)=(i/n^2,(i \% n^2)/n,(i \% n^2) \% n )$ and 
$(j_x,j_y,j_z)=(j/n^2,(j \% n^2)/n,(j \% n^2) \% n )$ of the lattice, 
where the divisions are considered in the integer number field and
 \% is intended as the division remainder operation. 
This naturally defines the distance of two elements of the matrix, $m_1=(i_0,j_0)$, $m_2=(i_1,j_1)$ as
the distance in periodic boundary conditions of the corresponding sites on the lattice; from this distance one can
then determine any $k$-th neighbor of a lattice site.

\paragraph*{$\mathbf{M}_2$ matrices (exponential).}
The random matrices have been obtained in a two step process; consider a $m \times m$ square matrix: 
first, the diagonal elements $D(1)$,$D(2)$,...,$D(m)$, have been determined with a uniform random number between $a + (b-a)/4$ and 
$b - (b-a)/4$ for each diagonal element. Then, each off-diagonal element is determined in the lower triangular part of the matrix;
the eigenvalues interval is controlled with the Gerschgorin circle theorem in the following way: if the element $(i,j)$ is $x$, then 
it will contribute to the Gerschgorin radii $R(i)$ and $R(j)$ of both the $i$ and $j$ row due to the symmetry of the matrix; more precisely, 
adding an off-diagonal element $x$ at position $(i,j)$ will increase both $R(i)$ and $R(j)$ by a positive quantity $|x|$. Fixing a row $i$, 
each element $j_0,j_1,...,j_{i-1}$ is a random number uniformly distributed between $-r/(i-j)$ and $r/(i-j)$, where 
$r = \min [D(i)-R(i)-a,b-D(i)-R(i),D(j)-R(j)-a,b-D(j)-R(j)]$; the division by $i-j$ ensures an almost complete filling of the matrix elements. 
This procedure guarantees that the eigenvalues of the resulting matrix lie in the interval $(a,b)$.

\bibliographystyle{unsrt}

\end{document}